\documentstyle[aps,epsfig]{revtex}
\newcommand{\ccn}[2]{C^{\phantom{\dagger}}_{\mathstrut#1\mathbf{#2}}}    
\newcommand{\ccc}[2]{C^{\dagger}_{\mathstrut#1\mathbf{#2}}}              

\begin{document}
\title{Superconductivity in the Model of Elastic Jelly}
\author{I.~M.~Yurin and V.~B.~Kalinin}
\maketitle
\address {\textit{Institute of Physical Chemistry, Leninskiy prosp. 31,
GSP-1, Moscow, 119991, Russia}}
\begin{abstract}
 In this work, a question is tackled concerning the formation of a
superconducting condensate in an earlier  proposed model of "elastic jelly",
in which phonons of the valent skeleton play the part of initiating ones.
It was shown that in distinction from the BCS theory, the momenta of forming
electron couples are different from zero. This fact changes the pattern of the
description of the superconductivity phenomenon in the proposed model. First,
the gap in the one-electron spectrum appears due to the effect of a
"mean field" on the energy of one-electron state from the side of
occupied states, the nearest neighbors over the momenta grid. Second,
the condensate is formed by one-electron states with energies below
that of the gap edge. This is why the Fermi-condensation arises in
the system. Third, in the proposed theory the electron couples appear
in the form of low-energy excitations, i.e., as those with the minimum
amount of energy per excitation electron. Hence, their role is minimized
to that of low-energy excitation with the minimum energy per electron,
and they are no more "bricks" the superconducting condensate is made of,
as the case is in the BCS theory.
\newline
PACS number(s): 74.20.Mn, 74.72.-h 
\end{abstract}
\section{Introduction}

 A considerable number of fundamental works 
have been devoted to tackling the superconductivity (SC) problem 
and, in  particular, a high temperature superconductivity one (HTSC).
The opinion on the present state of the problem 
may be formed considering the works ~\cite{bib-1,bib-2,bib-3}.
Most relevant publication are to some extent associated 
with the BCS theory~\cite{bib-4}. Now, for metals enumerated in
the Periodic System of elements (PSE), the superconductivity theory
is considered to be more or less completed and the validity of the
BCS theory leaves no room for doubt on the part of the majority of
authors. At the same time in ~\cite{bib-5,bib-6,bib-7,bib-8} a conclusion
was drawn that the BCS theory may be not the only possible way to explain
the phenomenon of SC.

 Outstanding results ~\cite{bib-9} obtained by Bednortz J.G. and Muller A.K.
in 1986 for YBa$_{2}$Cu$_{3}$O$_{7-\delta}$ caused the great sensation among
chemists, physisists and material researchers. Since that time the highest
value for $T_s$ at about 164K was observed in HgCa$_{2}$Ba$_{2}$Y$_{3}$Cu$_{8+\delta}$
ceramics. 

There are grounds available highlighting a possibility of observing high
$T_s$ in intercalates and nanotubes ~\cite{bib-3}. In our opinion, an
increase in the value of $T_s$ and the observation of HTSC are very
plausible in high-pressure (HP) phases. Almost twenty-five years ago,
SC was reported to have been observed in AuGa$_2$ ~\cite{bib-10}.
Comparatively high values of $T_s$ were observed in fullerenes,
HP phases with respect to graphite,
and their derivatives fullerides C$_{60}$M ~\cite{bib-11,bib-12}, where M stands
for K, Rb, Cs. The idea of obtaining HTSC in HP phases of Li$_{3}$P or
Li$_{3}$N is at present at the stage of being accepted ~\cite{bib-13,bib-14}.
In the last publications on this topic, SC is investigated in a HP phase of
MgB$_2$ ~\cite{bib-15,bib-16,bib-17}.
In conjunction with the problem under consideration, we also can't help
mentioning a hypothetical phase of metallic hydrogen ~\cite{bib-18} and
nitrogen (a private communication).

 The obtained results, especially the identification of HTSC with the
values of $T_{s}>100 K$, are rather inconsistent with both the predictions
of the BCS theory proper ~\cite{bib-19}, which had been made before
discovering HTSC, and any of its complimentary modifications, among
which the theory of bipolarons has to be singled out ~\cite{bib-2}.
In connection with this, we note that most HTSC materials
in stoichiometic phase are,
strictly speaking, dielectrics, whose conductivity is fully determined by
low concentrations of electroactive defects, whereas the theory
of bipolarons gives
low values of $T_s$ as the Fermi level approaches the edges of zones.
Thus, it does not explain HTSC for the aforementioned class
of compounds. 

In Russia, one usually considers the results of microscopic calculations
are in good agreement with the predictions of the BCS theory, including
the case of HTSC ~\cite{bib-20} as well. There is also another viewpoint
which is doubtful about the reliability of any fine calculation performed
by and large within the scope of the zone theory ~\cite{bib-21}. 

The problem is that a lot of poorly substantiated approximations did appear
for the years of the existence of the zone theory. Among these, for instance,
the procedure of accounting for the screening effects seems to be quite
harmless. These approximations were introduced by virtue of  quite relevant
reasons associated with an insufficient quick operation of computer
equipment to perform strict calculations of the zone structure of metals,
including that of the Periodic System. The application of these procedures,
however, renders the calculation of the electron-phonon interaction (EPI)
constants quite an unpredictable operation, depending only on theoretical
preferences of a specialist in the field of zone computations. 

From our viewpoint, the main source of difficulties consists in that all
the calculations associated with EPI pursue the goal of reducing the
Hamiltonian of the system to the Fr\"ohlich form. It has been known for
a long time ~\cite{bib-19} that the Fr\"ohlich model does not allow a
consistent accounting for the electron-electron interaction (EEI),
because the Coulomb interaction cannot still be reasonably separated
into that already accounted for in the initial values of Fr\"ohlich
Hamiltonian constants, and that which will thereafter show up in
the renormalization of the initial phonons owing to EPI.
Accordingly, all the microscopic  calculations of the EPI constants are,
strictly speaking, inconsistent.

Just these considerations underlied the need of a revision of the
fundamentals of the SC theory, which was undertaken in studies
~\cite{bib-22,bib-23,bib-24}. It was shown the attraction between
electrons exists in a long-wave range due to an exchange of virtual
phonons. This question was not discussed in earlier works (for
example, ~\cite{bib-25}) devoted to the problem being tackled.
We believe this was due to the following two basic causes.

First, in the framework of "procedure to account for the screening
effects" the function of the permittivity of electronic plasma was
incorrectly used for the removal of "undesirable" singularities
from the matrix elements of EPI in the long-wave limit.

 Second, the simplified form of the effective EEI used in study ~\cite{bib-25}
did not suppose substantial differences in the electron interaction
both in the vicinity of the Fermi surface and far off it. 

 The authors of ~\cite{bib-25} well understood the disadvantages of the
aforementioned approaches. Remind that in ~\cite{bib-19} serious
forebodings were caused just by these approximations, which gave
rise to a detailed discussion. Moreover, an idea was put forward
that in a further development of the "jelly" model one should
have refrained from the use of ion-plasma oscillations as
initiating phonons.

 The accounting for just these remarks underlied the investigation
in the long-wave range of EEI, which had been undertaken in our
studies ~\cite{bib-22,bib-23,bib-24}, where the oscillations of the valent
skeleton serve as initiating phonons. It was shown the
adequate inclusion  of the phonon--phonon interaction in the
examination process enables one to avoid the application of poorly
substantiated procedures to "account for the screening effects".
And finally, the EEI potential was calculated without any
limitations as to its form.

It was shown that the unitary transformation effected in the framework
of the model suggested in ~\cite{bib-23,bib-24} can reduce the Hamiltonian
of the electronic system of an indefinite monoatomic metal to the following
form ($\hbar=1$):
\begin{eqnarray}
 \tilde{H}_{\mathrm tot}
 &=& \sum_{\mu} \int E_{p}
     \ccc{\mu}{p}\ccn{\mu}{p}
      d{\mathbf p}
  + \tilde{H}_{ee}
  \label{eq-1} \\
 \tilde{H}_{ee}
 &=& \sum_{\mu}\sum_{\nu}\int\!\!\!\int\!\!\!\int 
     \tilde{U}({\bf p},{\bf k},{\bf q})
      \ccc{\mu}{p+q}\ccc{\nu}{k-q}
       \ccn{\nu}{k}\ccn{\mu}{p}
        d{\mathbf p} d{\mathbf k} d{\mathbf q}
 \label{eq-2}\ ,
\end{eqnarray}
and for $\tilde{U}({\bf p},{\bf k},{\bf q})$ at $p, k \approx K_F$ we have
\begin{equation}
 \tilde{U}({\bf p},{\bf k},{\bf q})
 =- 4 \left(\frac{zm}{3M}\right)^2
      \frac{G_{ee}}{q^2}
       \frac{K_F^2}{q^2-\chi_1}
        \frac{K_F^2}{q^2-\chi}
 \label{eq-3}
\end{equation}

where $E_{p}=\frac{p^2}{2m}$,
$G_{ee}=\frac{e^2}{4\pi^2\epsilon}$,
$e$ is the electron charge,
 $\ccc{\mu}{p}$ and $\ccn{\mu}{p}$
are operators of the creation and annihilation of electrons
with a momentum $\bf p$, respectively;
 $\mu$, $\nu$ are spin indices,
$\epsilon$ is the permittivity of the valent skeleton,
 $M$ is the mass of ion,
 $m$ is the mass of a zone electron,
 $z$ is the number of conductivity electrons per elementary cell;
$\chi=4\left(m^2\tilde{S}^2-(zm/3M)K_F^2\right)$,
$\chi_1=\left(1+(zm/3M)(K_F^2/\lambda^2)\right)\chi$,
$K_F$ is the value of Fermi vector, and $\tilde{S}$ is
the sound velocity. 

 The potential of the form ~(\ref{eq-3}) leads to an instability relative
to the formation of electron pairs. In distinction from the BCS theory,
the momenta of electron couples are different from zero, while the couples
bonding energy $E_b$ if a typical relationship $\frac{\chi}{2m}<<E_b$
for SC systems is fulfilled, may be estimated as follows:
\begin{equation}
  E_b \sim \left(\frac{zm}{M}\right)^2\frac{G_{ee}K_F^4}{Q^3}
\label{eq-4}
\end{equation} 
where $Q=\chi^{1/2}$.

 It is obvious that $E_b\sim M^{-1/2}$, and this is consistent with the
observations of an isotopic effect in a series of metals ~\cite{bib-26}.
We also note that the coherence length $l_{coh}\sim Q^{-1}$.

\section{SC condensate in the model of elastic jelly}
The further discussion will be carried out for the final system - a cubic
crystal $L\times L\times L$ with periodic boundary conditions.
 
 Consider in the framework of Eqs.~(\ref{eq-1})-~(\ref{eq-3}) a wave
function of the bound state of an electron pair with a minimum energy. Its
state will obviously be s-type. With increasing $\tilde{S}$, its wave function
will first become hydrogen-like, and then delta-like, so that
at $\tilde{S} \to \infty$ it proves to be possible to derive for
the state $\left|{\bf R}\right\rangle$ of an electron couple with a momentum
${\bf R}$ at an accuracy of up to the normalizing factor the following expression:
 \begin{equation}
\left|{\bf R}\right\rangle=\sum_{{\bf p}+{\bf k}=
{\bf R}}\tilde\delta({\bf k}-{\bf p})
C^+_{\uparrow {\bf p}}C^+_{\downarrow{\bf k}}\left|0\right\rangle
\label{eq-5}
\end{equation}
where $\tilde\delta({\bf p}-{\bf k})$ is a discrete delta-shaped function,
which is determined as follows: 
\begin{equation}
\tilde\delta({\bf p}-{\bf k})=\left\{\begin{array}{l}
1, \quad \mbox{if}\quad|{\bf p_{\alpha}}-{\bf k{_\alpha}}|\le \frac{2\pi}{L}, \quad \alpha=x,y,z\\
0\mbox{ in the rest cases.}\\
\end{array}
\right.
\label{eq-6}
\end{equation}

 Hence, it follows that in finite crystal, at least in the limit under consideration
$\tilde{S} \to \infty$, the occupation numbers remain good quantum numbers in
describing the states with the maximum bonding energy. A possibility arises to
account for these states in the Fermi-liquid theory by introducing a delta-shaped
correction to the Landau function. For the sake of simplification, we assume its
regular part may be accounted for by introducing a weak temperature dependence
of the electron mass. The Hamiltonian of the system will then become
\begin{equation}
H=
\sum_{\mu, {\bf p}}
E_{p}\hat{n}_{\mu {\bf p}}-\frac{\Delta}{27}\sum_{{\bf p},{\bf k}}
\tilde\delta({\bf p}-{\bf k})
(\hat{n}_{\uparrow {\bf p}}\hat{n}_{\downarrow{\bf k}}+\hat{n}_{\downarrow {\bf p}}\hat{n}_{\uparrow{\bf k}})
\label{eq-7}
\end{equation}
where $\hat{n}$ are the operators of the occupation numbers,
$\Delta=\frac{27}{2}E_{b}$. The Hamiltonian so written is able of correctly
describing both the statistics and the energy spectrum of one- and two-particle
excitations. 

 The existence of the bound multiparticle states of electrons, for which, like for
the considered s-state as well, the occupation  numbers remain to be good quantum
numbers in the limit $\tilde{S} \to \infty$, is accounted for by introducing the
dependence of $\Delta$ on the occupation density $\tilde\rho$:
\begin{equation}
\Delta(\tilde\rho)=(1-\beta)\Delta_0+4\beta(\tilde\rho-1/2)^{2}\Delta_0
\label{eq-8}
\end{equation}
In Eq.~(\ref{eq-8}), the symmetry of the bonding energy between electrons
with $k>K_F$ and holes with $k<K_F$,  which is rather approximate in real systems,
has been accounted for.

 In the mean field approximation, the Fermi-Dirac distribution function is
expressed through the local energy of a quasi-particle $\tilde{E}$, and a
self-consistency equation for the occupation density is as follows ($k_{B}=1$):
\begin{figure}[thb]
$$\psfig{figure=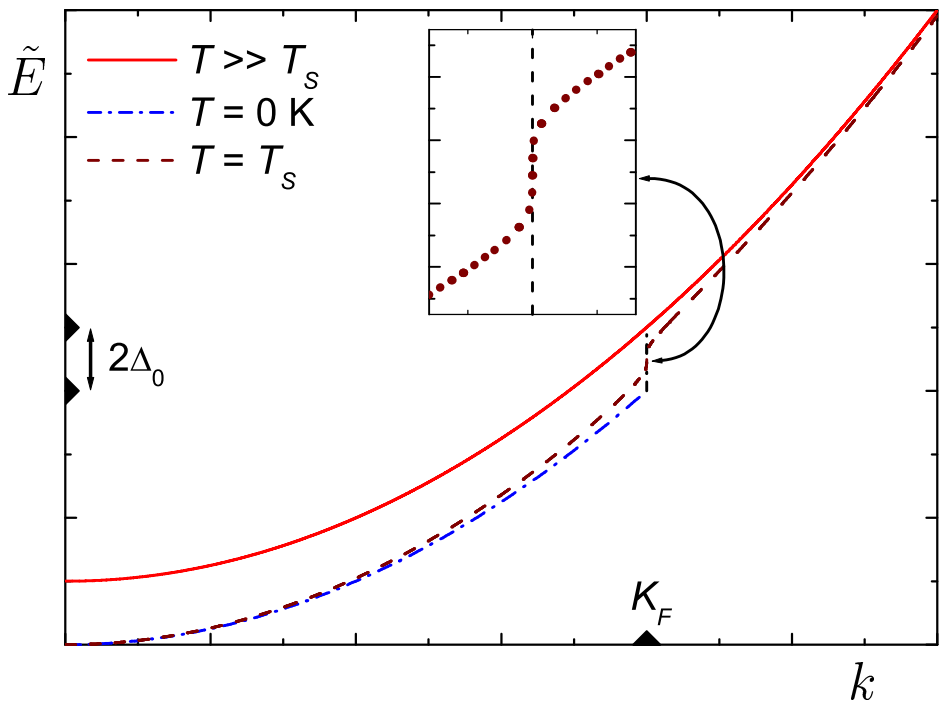,height=100mm}$$
   \caption{Temperature dependence of one-electron spectrum
in the model of elastic jelly (schematic)}  
\end{figure}
$$
\tilde\rho=\left[1+
exp\left(\frac{\tilde{E}-\xi}{T}\right)\right]^{-1}
$$
\begin{equation}
\tilde{E}=E-2\tilde\rho\Delta(\tilde\rho)
\label{eq-9}
\end{equation}

 The evolution of a one-electron spectrum of the system with its temperature
varying becomes obvious from Fig. 1. At high temperatures, this spectrum is
described by an ordinary parabola; and at $T=T_s$ the spectrum exhibits a
peculiar feature which, at $T<T_s$, transforms into a gap. 

 The transition temperature $T_s$ may be estimated by the equation
$\partial\tilde\rho/\partial{E}|_{\tilde{E}=\xi} = -\infty$.
Then we derive for $T_s$
\begin{equation}
T_s=\Delta_{0}\frac{1-\beta}{2}
\label{eq-10}
\end{equation}

 So, the SC condensate is formed by electrons with the
energy $\tilde{E}<\xi$ at $T<T_s$. The couples of electrons with momenta
$k>2K_F$ or holes with $k<2K_F$ in a superconducting condensate prove
to be the lowest-energy excitations from the viewpoint of the excitation
energy per particle. 

 The parameters of the theory may be determined in Landau's style by
comparison with the experiment. As a matter of fact, for the threshold
voltage of a normal metal-dielectric-superconductor tunnel transition,
corresponding to one-electron processes, we have the following equation:
\begin{equation}
eV_t=\Delta_0
\label{eq-11}
\end{equation}
which, together with Eq.~(\ref{eq-10}), forms a complete system
of equations to determine the parameters of the simplest version
of the theory. 
 The remaining bound multielectron states, for which the occupation numbers
are not good quantum numbers in the aforementioned sense, may be accounted
for by inserting the  nondiagonal terms in the Hamiltonian. Yet this does
not substantially distort the microscopic picture of the phenomenon by virtue
of two interrelated causes. Firstly, these terms are small in their absolute
values calculated per excitation electron, compared with the bound energy of
s-state, which has already been accounted for. Secondly, both the state of
the condensate itself and the state of excitations of electron or hole pairs
are separated by an energy gap from other states, into which they could have
passed under the effect of the nondiagonal terms being considered. Therefore,
in considering both the state of condensate and that of excitation of electron
(hole) couples, the arisen nondiagonal terms may be accounted for in the
framework of the perturbation theory without any substantial changes in
the wave functions and excitation spectrum.

 Let consider the aforesaid limit $\tilde{S} \to \infty$. The energies of the bound
states of the starting Hamiltonian ~(\ref{eq-1}) are of an analytical character
in relation to the parameter $\tilde S$ throughout the whole range of
admissible values. This is why the analyticity of the energy parameters of the
Fermi-liquid theory of Landau relative to the same parameter cannot raise
doubts. Then the presence of the aforementioned delta-shaped correction
to the Landau function for all the admissible values of $\tilde{S}$ seems
to be obvious.
 
 In conclusion, we note that the contribution from the Coulomb EEI, though
weakened by accounting for the correlation effects in a degenerated electron
plasma ~\cite{bib-24}, is able to destroy the distinctly formulated microscopic
picture of SC. However, the question being discussed is beyond the scope of the
posed task. 
 
\section{The condition for the onset of HTSC}
 A particular interest is spurred by the divergence of the bonding energy
$E_b$ at $Q \to 0$, following from Eq.~(\ref{eq-4}). In real systems,
this divergence should not reveal itself in the observed $T_s$ by virtue
of symmetry reasons  put forward in ~\cite{bib-23}. Simultaneously, with
varying parameters of specimens, there should be observed a maximum of the
transition temperatures if the following condition were fulfilled:
\begin{equation}
 \tilde{S}^2 \approx \frac{zm}{3M} V_F^2
\label{eq-12}
\end{equation} 
This condition can be with a satisfactory accuracy fulfilled in systems
with variable physical parameters, in which, provided Eq. ~(\ref{eq-12}) is
obeyed, HTSC should be observed.
  
 Let consider a semiconductor with a rather low concentration of electroactive
defects, undergoing a phase transition with a change in volume. In both phases,
the material is stable, and, as a rule,
$\tilde S^2=\partial P/{\partial\rho}>>\frac{zm}{3M} V_F^2$
(here $\rho=M/\Omega$). In unstable phases $\partial P/{\partial\rho}<0$.
It is possible to assume HTSC is observed in nonequilibrium systems, where
the phase transition is decelerated
$\tilde S^{2}=\partial P/{\partial\rho}\approx\frac{zm}{3M} V_F^{2}>0$
(the diamond-graphite transition is not realized exclusively due to
kinetic reasons). 

\section{Conclusions}
 The basic qualitative differences of the BCS theory from the suggested
SC model are listed below.

1. In accordance with the BCS theory, the superconducting condensate is
formed by the couples of electrons with zero momenta, and is therefore called
Bose condensate. 

In the proposed theory, the superconducting condensate is formed by electrons,
and, accordingly, transforms into the Fermi condensate.
 
2. In accordance with the BCS theory, the most low-energy excitations are
essentially the electron couples with zero momenta.

In the proposed theory, such excitations at the lowest temperatures are
the electron couples with momenta  $k>2K_F$ and hole couples with $k<2K_F$.

3. In the BCS theory, the ground state of the system is obtained through
the superposition of the states with different charges. It is this property
that underlies one of the formulations of the BCS theory using correlators
$<C^{+}C^{+}>$ and $<CC>$ that are different from zero. 

Now, in the proposed theory the corresponding correlators are strictly
equal to zero.
 
4. It is easy to show each of the states with a fixed charge, forming the
superposition of the ground state of the BCS theory, is stationary state,
and, hence, also ground one in conformity with the charge-conservation law.
Thus, a conclusion may be drawn on the eventual degeneration of the ground
state of the system in the BCS theory. This means its chemical potential
$\xi=\delta$$E/$$\delta$$n$ at $T=0 K$ must be strictly equal to zero;
therefore, the work function of electrons in SC also must be equal to zero.

 Now, in the proposed model such a requirement is irrelevant.

 The considered theories give different expressions for such parameters as
energy gap and coherence length. It is curious that the coherence length in
the BCS theory decreases as the bonding energy of a couple increases,
reaching a value on the order of several $\AA$ngstr\"oms~\cite{bib-1}. This
appears very strange in view of the Coulomb repulsion energy arising at such
a spacing.

 In the proposed model, the coherence length cannot be shorter than the value
of $(2m\tilde{S})^{-1}$. 

 Although one has achieved in the framework of BCS theory consistency of
the experimental data with the microscopic calculation results in (HT)SC
materials, this can hardly be considered a convincing proof in its favor.
A scepsis about this point should be paid heed to, being expressed by some
authors with regard to the reliability of the calculation of so subtle values
as EPI constants to be used in assessing the BCS theory.

 We also can't help noting here the absence of unitary transformations
linking Fr\"ohlich's Hamiltonian, in the framework of which one tries to
substantiate the introduction of Cooper couples, to any of the Hamiltonians
used in the BCS theory. In view of studies ~\cite{bib-22,bib-23,bib-24},
this fact should cause some sort of uneasiness.

 The interpretations of experimental results in both theories are fairly close,
differing merely in some details. In fact, the charges of low-energy
excitations of the system of couples of electrons and holes are equal
in the theories under examination. This is why the consequences of
Ginzburg-Landau theory will not differ substantially.

 Now, if the phenomena were investigated, being associated with the
dissipation of energy in an electronic system, their interpretation is
connected, as a rule, with the energy densities of excited states. By
and large, their form in both theories exhibits no qualitative
differences either. 

 Therefore, both the fulfilment of relationship ~(\ref{eq-12}) in observing
HTSC and the investigation of the work function of electrons in SC materials
so far seem to be sole practicable methods for the experimental comparison of
the theories. 

 One may expect the fulfilment of relationship~(\ref{eq-12}) with good
accuracy in the vicinity of the boundaries of stability of HP phases.
From this standpoint, it seems to be very promising to investigate the
kinetics of the phase transitions at HP with a view to searching for new
HTSC materials.

\section{Acknowledgements}
 We are grateful to Dr. N.V. Klassen, Dr. V.V. Gromov, and Prof. Allan Solomon
for their interest of our work and support of it.


\end{document}